\documentclass{ecai} 
\usepackage{balance}
\usepackage{latexsym}
\usepackage{amssymb}
\usepackage{amsmath}
\usepackage{amsthm}
\usepackage{booktabs}
\usepackage{enumitem}
\usepackage{graphicx}
\usepackage{color}
\newcommand{\para}[1]{\vspace{1mm}\noindent\textbf{#1}}
\usepackage{listings}
\usepackage{xcolor} % For defining custom colors
\usepackage{subcaption}
\usepackage[export]{adjustbox}

\definecolor{codegreen}{rgb}{0,0.6,0}
\definecolor{codegray}{rgb}{0.5,0.5,0.5}
\definecolor{codepurple}{rgb}{0.58,0,0.82}
\definecolor{backcolour}{rgb}{0.95,0.95,0.92}

\lstdefinestyle{mystyle}{
    backgroundcolor=\color{backcolour},
    commentstyle=\color{codegreen},
    keywordstyle=\color{magenta},
    numberstyle=\tiny\color{codegray},
    stringstyle=\color{codepurple},
    basicstyle=\footnotesize\ttfamily,
    breakatwhitespace=false,
    breaklines=true,
    captionpos=b,
    keepspaces=true,
    numbers=left,
    numbersep=5pt,
    showspaces=false,
    showstringspaces=false,
    showtabs=false,
    tabsize=2,
    xleftmargin=10pt,
    xrightmargin=10pt,
    framexleftmargin=5pt,
    framexrightmargin=5pt,
    framextopmargin=5pt,
    framexbottommargin=5pt,
}

\newcommand{\BibTeX}{B\kern-.05em{\sc i\kern-.025em b}\kern-.08em\TeX}

\begin{document}

%%%%%%%%%%%%%%%%%%%%%%%%%%%%%%%%%%%%%%%%%%%%%%%%%%%%%%%%%%%%%%%%%%%%%%%%

\begin{frontmatter}

%%% Use this command to specify your submission number.
%%% In doubleblind mode, it will be printed on the first page.

\paperid{123} 

%%% Use this command to specify the title of your paper.

\title{AI-Powered Immersive Assistance for Interactive \\
Task Execution in Industrial Environments}

%%% Use this combinations of commands to specify all authors of your 
%%% paper. Use \fnms{} and \snm{} to indicate everyone's first names 
%%% and surname. This will help the publisher with indexing the 
%%% proceedings. Please use a reasonable approximation in case your 
%%% name does not neatly split into "first names" and "surname".
%%% Specifying your ORCID digital identifier is optional. 
%%% Use the \thanks{} command to indicate one or more corresponding 
%%% authors and their email address(es). If so desired, you can specify
%%% author contributions using the \footnote{} command.

\author[A]{\fnms{Tomislav}~\snm{Duricic}\thanks{Corresponding Author. Email: tduricic@know-center.at.}}%\footnote{Equal contribution.}}
\author[A]{\fnms{Peter}~\snm{Müllner}} 
\author[A]{\fnms{Nicole}~\snm{Weidinger}}%\footnotemark}
\author[A]{\fnms{Neven}~\snm{ElSayed}} 
\author[A,B]{\fnms{Dominik}~\snm{Kowald}} 
\author[A,B]{\fnms{Eduardo}~\snm{Veas}} 

%\author[A]{\fnms{Tomislav}~\snm{Duricic}\orcid{0000-0002-7229-9596}\thanks{Corresponding Author. Email: tduricic@know-center.at.}}%\footnote{Equal contribution.}}
%\author[A]{\fnms{Peter}~\snm{Müllner}\orcid{0000-0001-6581-1945}} 
%\author[A]{\fnms{Nicole}~\snm{Weidinger}\orcid{0009-0003-7265-3668}}%\footnotemark}
%\author[A]{\fnms{Dominik}~\snm{Kowald}\orcid{0000-0003-3230-6234}} 
%\author[A]{\fnms{Neven}~\snm{ElSayed}\orcid{0000-0002-5153-8084}} 
%\author[A,B]{\fnms{Eduardo}~\snm{Veas}\orcid{0000-0002-0356-4034}} 

\address[A]{Know-Center GmbH}
\address[B]{Graz University of Technology}

%%% Use this environment to include an abstract of your paper.

\begin{abstract} 
Many industrial sectors rely on well-trained employees that are able to operate complex machinery. 
In this work, we demonstrate an AI-powered immersive assistance system that supports users in performing complex tasks in industrial environments. 
Specifically, our system leverages a VR environment that resembles a juice mixer setup.
This digital twin of a physical setup simulates complex industrial machinery used to mix preparations or liquids (e.g., similar to the pharmaceutical industry) and includes various containers, sensors, pumps, and flow controllers. 
This setup demonstrates our system's capabilities in a controlled environment while acting as a proof-of-concept for broader industrial applications. 
The core components of our multimodal AI assistant are a large language model and a speech-to-text model that process a video and audio recording of an expert performing the task in a VR environment. 
The video and speech input extracted from the expert's video enables it to provide step-by-step guidance to support users in executing complex tasks.
This demonstration showcases the potential of our AI-powered assistant to reduce cognitive load, increase productivity, and enhance safety in industrial environments. 
\end{abstract}

\end{frontmatter}

%%%%%%%%%%%%%%%%%%%%%%%%%%%%%%%%%%%%%%%%%%%%%%%%%%%%%%%%%%%%%%%%%%%%%%%%

\section{Introduction}

As the industrial sector continues to embrace technological advancements, integrating Artificial Intelligence (AI) into operational processes has become a key driver of efficiency, safety, and innovation~\cite{shaji2022artificial}. 
In this vein, this paper introduces an AI assistant designed for immersive training, leveraging the synergies of multimodal AI and Virtual Reality (VR) technology to support task execution within industrial environments. The motivation for such tools arises from the increasing complexity of industrial machinery, which burdens operators with a cognitive load that can compromise both productivity and safety~\cite{carvalho2020cognitive}. Additionally, there is a need to improve machine operator training and adaptability in the face of evolving industrial standards and practices, while also providing support in situations where a knowledgeable expert is unavailable~\cite{patalas2019approach}.

Furthermore, additional challenges include the unavailability of physical machinery for training due to cost, the infrequent nature of certain tasks performed by experts only during assembly, and the significant need for upskilling in an ever-changing job market~\cite{collino2022reducing}. These challenges underscore the importance of creating a flexible and comprehensive virtual solution, allowing trainees to experience key activities in a safe, immersive environment~\cite{radhakrishnan2021systematic}.

In response, our approach, showcased on a virtual juice mixer testbed that is a digital twin~\cite{sjarov2020digital} of an actual physical setup
%\footnote{A digital twin is a virtual replica of a physical device used to run simulations before actual devices are built and deployed. Our virtual juice mixer is an exact digital counterpart of a physical juice mixer.}%, allowing us to simulate and analyze operations without the risks and costs associated with physical testing.}
, aims to demonstrate how AI assistants can offer a scalable and effective solution to these challenges and enhance interactive task execution across a wide array of industrial applications.

The novelty of our approach lies in deploying an interactive AI assistant powered by a large language model (LLM) that uses audio transcripts to dynamically generate step-by-step guidance for immersive and intuitive training. These transcripts are extracted from a video of an expert performing the task in a VR environment and serve as the primary context for guidance. The virtual testbed replicates the setup of its physical counterpart, ensuring that our simulations and training scenarios align with real-world operations~\cite{jiang2021industrial}. The LLM-based assistant processes both text and speech inputs, dynamically adapting outputs to address user needs at each step.

By implementing this system on a VR platform, we demonstrate the practical application of our AI assistant in simplifying complex industrial tasks and its potential to improve operational efficiency and learning effectiveness. This paper details the implementation and use of our assistant, illustrating how it integrates with VR to provide immersive, intuitive support for industrial operations. Through this exploration, we contribute to the discourse on AI's role in industrial automation, offering insights into its potential to improve interactions with complex machinery. In the next section, we outline the challenges of integrating immersive technologies into industrial operations and the role of AI in enhancing safety and efficiency.

\section{Background}
%\textbf{TODO: background and challenges and the juicemixer use-case description which we can shift then, we need this motivation to explain the people why we need the recommender system}
\para{Industrial Immersive Environments.} The integration of immersive technologies, such as digital twins and VR, into industrial settings represents a paradigm shift in how operations and training are conducted. Digital twins offer a digital representation of physical systems, enabling real-time monitoring, simulation, and control of industrial processes without direct physical interaction~\cite{jiang2021industrial,stavropoulos2022digital}. Simultaneously, VR has emerged as a crucial tool for immersive training, allowing operators to experience and interact with complex machinery in a safe, virtual environment before applying these skills in the real world~\cite{hasan2017virtual,randeniya2019virtual,gavish2015evaluating}. These technologies have not only streamlined operational procedures but also significantly minimized risks, contributing to a safer and more efficient industrial environment~\cite{voinea2023mapping,babalola2023systematic}.

\para{Challenges in Industrial Operations and the Role of AI.} Despite advancements in immersive technologies, industrial operations continue to face significant challenges. Increasing complexity of machinery and rapid technological and regulatory changes demand expertise and flexibility from operators~\cite{russmann2015industry,sahoo2022smart,alkan2018complexity}. These challenges, coupled with the potential for human error under high cognitive load, underscore the need for innovative solutions to support operators in real-time decision-making and task execution. Moreover, the potential unavailability of experts, due to distance or scheduling conflicts, further complicates these challenges, underscoring the importance of an autonomous guidance system~\cite{torres2021classification,carvalho2020cognitive,naef2022decision}. Our goal is to enable trainees to access prerecorded information contextualized to their needs on the fly. Notable attempts in the past relied on continuous tracking of visual attention, coupled with the recognition of focused objects, to retrieve video snippets~\cite{leelasawassuk2017automated}. Another attempt introduced a new benchmark dataset and explored the use of foundation models to address similar challenges~\cite{bao2023can}. 

AI has emerged as a key enabler in overcoming these obstacles by augmenting human capabilities with intelligent, context-aware assistance. Leveraging AI, industries can create systems capable of analyzing complex data to offer predictive insights, automate routine tasks, and provide adaptive, step-by-step guidance tailored to the operator's current task and environment~\cite{lee2020industrial,javaid2022artificial}. The fusion of AI with immersive technologies paves the way for a new generation of assistance systems that are more intuitive, interactive, and capable of significantly reducing the cognitive load on operators, thus mitigating the risks associated with complex industrial operations~\cite{tiple2024ai,chheang2024towards}.

This evolving landscape of industrial settings, coupled with the transformative capabilities of AI, lays the groundwork for building and showcasing our system. Our approach goes beyond mere recognition of actual context and allows trainees to pose queries and interact with the content guided by a multimodal AI assistant. %Our AI-powered immersive assistance system is designed to address these challenges, offering a novel approach to interactive task execution that enhances safety, efficiency, and adaptability in industrial operations.

%%%%%%%%%%%%%%%%%%%%%%%%%%%%%%%%%%%%%%%%%%%%%%%%%%%%%%%%%%%%%%%%%%%%%%%%

\section{Demo Setup}
%\textbf{TODO: Digital twin question? AI assistant for VR digital twin, no need to say we have  aphyisical one, but mention that it is a digital twin, another way would be to say industrial setting, cite a paper for a digital twin, another way would be to add a footnote and explain it in one sentence, physical machine with real-time sensor data streaming, early next week discuss with Eduardo and Nicole}

%\textbf{TODO: create an Overleaf about 4 pillars: GUI, Authoring, Multimodal user context, Knowledge representation/prompting}
%\textbf{TODO: GUI for example distributed vs panel, situated recommending system (recommending interaction coming from the recommender system) vs window recommending system}
%\textbf{TODO: connect the solution with the problem in AR, another way is HCI, i.e., the interaction with the user, the usability of the technique, how the people can use it}
%\textbf{TODO: idea description, expected contribution, research question}
%\textbf{TODO: read upon the GUI current research and findings, start with the VR anatomy paper}

%\textbf{TODO: how the demo setup is going to look like at the conference, describe the devices, environment and what will be shown in the demo}
The live demonstration showcases our AI-powered immersive assistance system in VR. Users experience an interactive setup featuring the virtual juice mixer testbed, designed to simulate a complex industrial machine with containers, sensors, pumps, and flow controllers. The demo provides participants with an immersive experience that highlights the AI assistant's capabilities. The video for the demo is hosted on YouTube and is available at [\url{https://www.youtube.com/watch?v=iFdK_TUcVQs}].

\para{Development Framework.} The system is developed using Unity\footnote{\url{https://unity.com/}} and Oculus VR\footnote{\url{https://developer.oculus.com/}}, with Meta Quest\footnote{\url{https://www.meta.com/at/en/quest/products/quest-2/}} serving as the primary device for the demonstration. The development process involves creating an environment that accurately replicates the juice mixing operation, allowing users to interact with virtual components and understand the task's operational principles.

\begin{figure}
\centering
\includegraphics[width=0.8\linewidth]{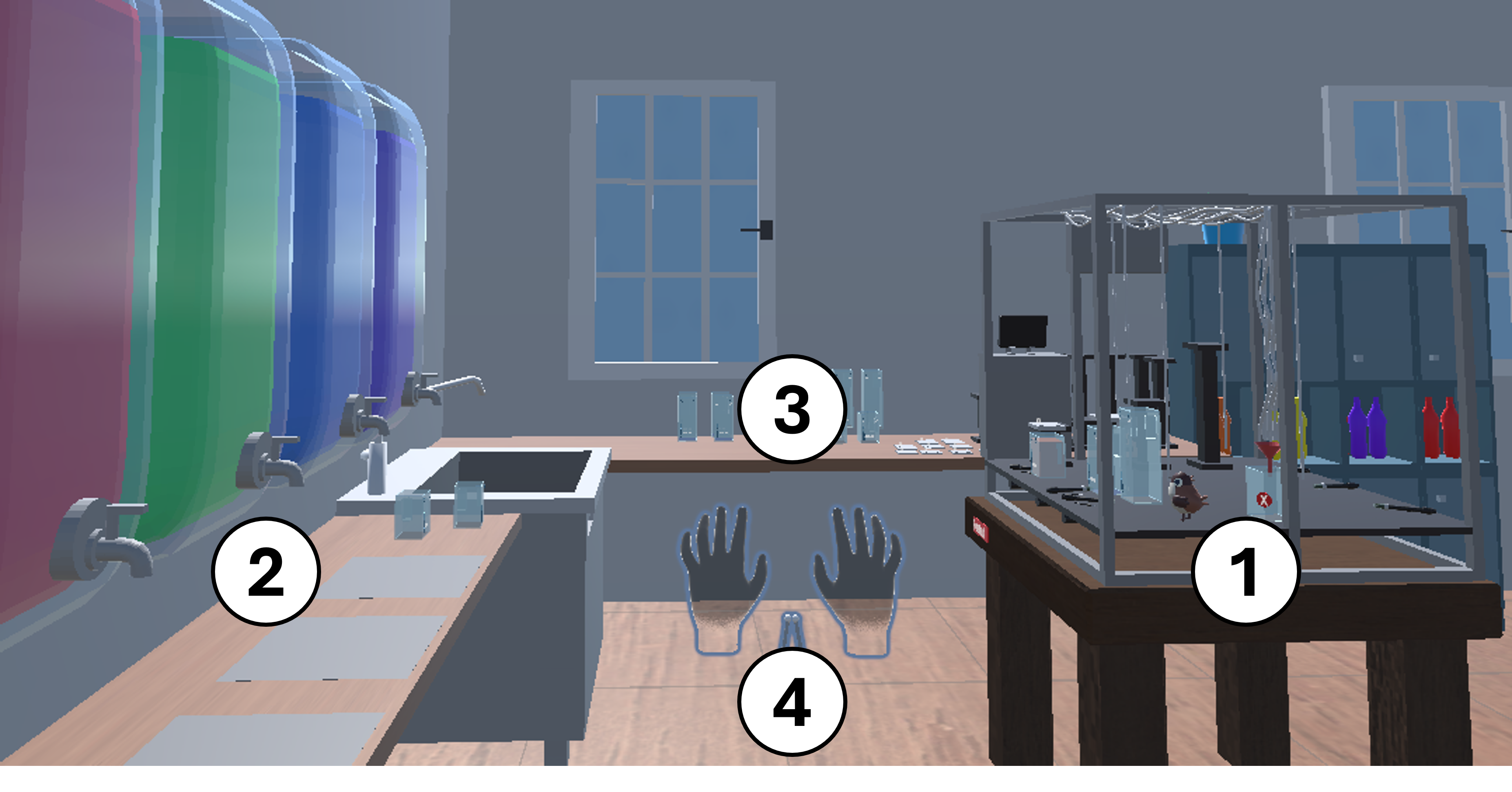}
\caption{Overview of the virtual juice mixing setup in VR. Key components are highlighted: (1) Juice Mixer, (2) Juice Station, (3) Spare Part Station, and (4) Controller/Hands as input, which illustrates the user interaction within the immersive environment.\vspace{4mm}}
\label{fig:juicemixer}
\end{figure}

\para{Juice Mixer Digital Twin.} In our VR setup, the juice mixer, juice station, and spare part station form the core of the interactive environment simulating the juice mixing process (see Figure \ref{fig:juicemixer}). The juice mixer resembles a machine used in pharmaceutical and chemical domains. This setup allows users to interact with digital twin, helping them grasp the operational principles and functionalities of the juice mixing operation in an immersive manner.

\para{Operational Task Flow.}
The task flow is structured to guide users through the juice mixing process in a sequential and logical manner, utilizing VR controls for interaction with the virtual equipment:

\begin{itemize}
\item \textit{Preparation:} Users select and pick up a container, placing it under a spout at the juice station, where the container is automatically filled with their chosen type of juice. A visual indicator shows the fill level of the container.
\item \textit{Assembly:} Once filled, users attach the lid and relevant sensors (temperature and pH sensors) to the container. At this stage, they also connect a tube from the pump to the container, enabling the forthcoming mixing process. These components are designed for easy attachment through intuitive controller actions, enhancing the realism of the simulation.
\item \textit{Mixing:} With the setup complete, the user proceeds to the mixing stage, interacting with virtual knobs to adjust the pump's strength and operational mode. The process provides hands-on experience in managing the mixing intensity and duration, closely replicating the actual operational controls.
\item \textit{Final Steps:} After mixing, users examine the final mixture, assessing the outcome of their efforts. This step not only concludes the task flow but also reinforces the learning objectives by enabling users to directly observe the results of their actions.
\end{itemize}

This simulation provides users with a comprehensive understanding of the juice mixing process within a controlled, risk-free virtual environment. The interactive setup enhances training efficacy, allowing operators to master complex machinery operations without the physical risks typically associated with industrial environments.

%%%%%%%%%%%%%%%%%%%%%%%%%%%%%%%%%%%%%%%%%%%%%%%%%%%%%%%%%%%%%%%%%%%%%%%%

\section{AI-Powered Immersive Assistance}

\begin{figure*}[t!]
  \centering 
  \includegraphics[width=0.9\linewidth]{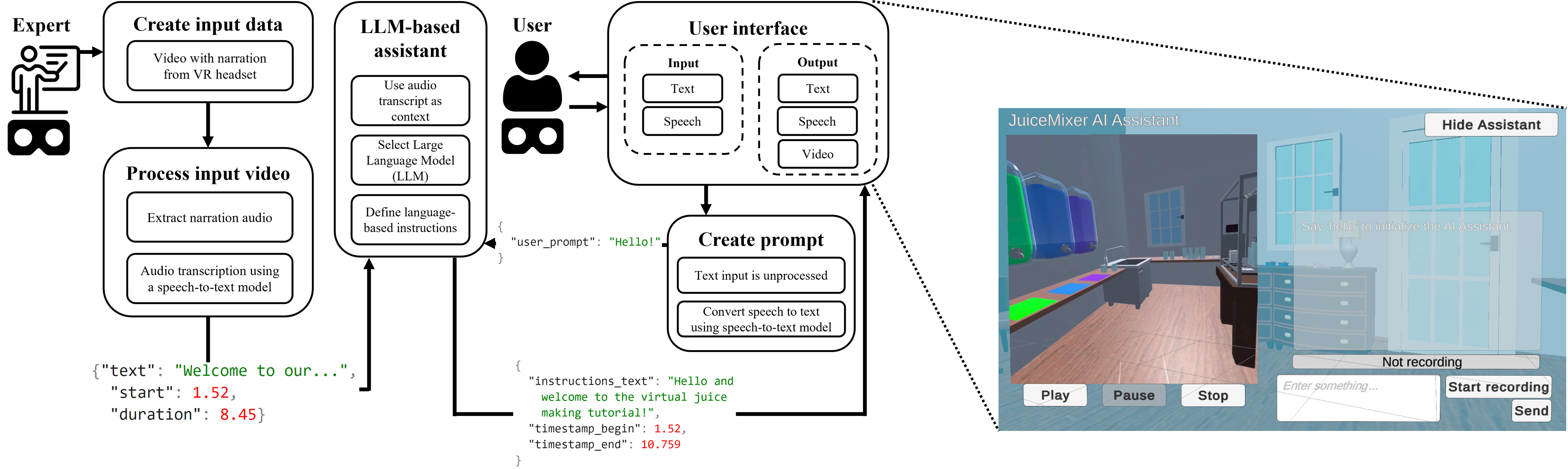}  
  \caption{System-level (left) and user-level (right) perspective of the immersive AI assistant.
  The assistant needs an expert to perform the task, and the expert's narration is transcribed to text, which serves as  context for the LLM.
  Given this context and text or speech input from the user, the LLM generates multimodal instructions that guide the user through the task. 
  These instructions are presented to the user within a VR environment with media controls, text command input, and voice interaction to facilitate user engagement with the AI Assistant.
  \vspace{2mm}}
  \label{fig:architecture-gui} 
\end{figure*}

%The AI assistant's purpose is to support juice mixer operation training in an immersive and interactive way. We use an expert video with narration as input, which the system processes to setup an interactive assistant. This method allows individuals to learn at their own pace and receive expert advice indirectly, enhancing the training experience in scenarios where direct expert interaction is unavailable. Next, we delve into the implementation details of the AI assistant and user interactions within the VR environment.

The AI assistant supports immersive and interactive juice mixer operation training. It uses a narrated expert video as input to guide trainees through an interactive assistant, allowing learning at their own pace when direct expert interaction is unavailable. Next, we delve into the implementation details (as depicted in Figure~\ref{fig:architecture-gui}) of the AI assistant and user interactions within the VR environment.

\para{Expert Video Creation and Processing.} The development of our AI assistant for machine operation training starts with capturing a video of an expert performing the task in the VR environment. The expert narrates and explains their actions step by step during the task. This narration is essential for capturing detailed instructions and insights for learning. After recording, the audio is transcribed into text using the OpenAI speech-to-text model\footnote{\url{https://platform.openai.com/docs/guides/speech-to-text}}, with timestamps included to preserve sequence information. This transcript is then converted into a JSON format, serving as the primary input for generating the AI assistant's instructional content.

\para{Creating an LLM-Based Assistant.} Using the OpenAI Assistants API\footnote{\url{https://platform.openai.com/assistants/}}, we employ the GPT-4 language model to power our AI assistant, which enhances the user experience by allowing interactive and intuitive communication. The transcript, already formatted in JSON from the expert's narrated video, provides a rich context that the LLM uses to guide users through the juice mixing process in the VR setting. This approach enables us to capture the expert’s knowledge effectively while simplifying the user's interaction with the system, enabling them to ask questions and receive instructions that are contextually aware and precisely timed.

\para{Defining AI Assistant Behavior and Communication.}
The AI assistant's behavior and communication style is simply defined by a set of explicit instructions using natural language within the OpenAI Assistants platform. These instructions dictate that the assistant's role is to guide users through the juice mixer operation in VR, step by step. The assistant uses a detailed transcript as context, timestamped and formatted in JSON, derived from an expert's video tutorial. The AI assistant is instructed with the following primary functions: \textit{(i) Guide Users -} Present and sequentially navigate through the juice-making steps, prompting users to confirm completion before proceeding. \textit{(ii) Respond to Queries -} Address user queries by referencing specific parts of the transcript, using timestamps to provide contextual accuracy. \textit{(iii) Troubleshoot Issues -} Offer solutions for common operational challenges as outlined in the transcript.

The assistant facilitates effective communication, ensuring each user gains practical skills and deep understanding of the juice mixing process. Initially, the assistant introduces itself, outlining its role and explaining how it assists in the juice-making process. It then continues guiding the user, responding to queries and providing detailed instructions based on the structured content of the expert’s narration.

%Responses from the AI are formatted in JSON, which enhances the clarity of the instructions and %aids in tracking interactions. For instance, a typical response might look like this:
%\begin{lstlisting}[basicstyle=\footnotesize\ttfamily, numbers=none]
%{
%  "instructions_text": "Please place the container under the juice outlet.",
%  "timestamp_begin": 2.3,
%  "timestamp_end": 5.4
%}
%\end{lstlisting}

Each response provides clear and detailed instructions for the current task or query and includes precise timestamps that dictate the playback window of the expert's video in the user interface. This targeted video playback visually highlights the specific step being discussed, enriching the learning experience by synchronizing instructional content with relevant visual cues. The assistant operates without external knowledge, relying entirely on the expert's video content to ensure a smooth and effective training experience.

\para{Interacting with the AI Assistant} The user interface for engaging with the AI assistant is designed to be both intuitive and user-friendly. Positioned next to the virtual juice mixer within the VR environment, the interface includes a dedicated panel that hosts several essential components for interaction, illustrated in Figure \ref{fig:architecture-gui}:

\vspace{-2mm}
\begin{itemize}
    \item \textit{Input Textbox:} Allows users to type their prompts, facilitating textual communication with the AI assistant.
    \item \textit{Audio Input Option:} Enables speech input, with recordings transcribed to text via OpenAI's speech-to-text model\footnote{\url{https://platform.openai.com/docs/guides/speech-to-text}}. Transcriptions appear in the input textbox for review or editing.
    %, and recordings automatically terminate after 10 seconds or can be ended by user command.
    \item \textit{Response Display and Audio Output:} After query submission, the AI assistant processes the prompt and displays the response in an output textbox. Simultaneously, the response is converted from text to speech\footnote{\url{https://platform.openai.com/docs/guides/text-to-speech}}, providing audio feedback.
    %, which is adaptable to different user environments and needs.
    \item \textit{Video Panel Integration:} The video panel displays clips from the expert video based on the AI assistant's timestamped responses, visually demonstrating the specific steps being discussed. 
    %This feature enhances understanding by aligning visual cues with verbal and textual instructions.
\end{itemize}
\vspace{-2mm}
The multimodal interface allows for flexible user interaction with the AI assistant, utilizing text, audio, and video outputs. The integration of these components ensures that all users can effectively navigate and master the juice mixing process within the VR environment, regardless of their specific learning needs or environmental conditions.

%\begin{figure}[t!]
%\centering
%\includegraphics[width=6cm]{architecture.png}
%\caption{Overview of the System Architecture: The process begins with a video tutorial from which a transcript with timestamps is extracted. This transcript, along with step-by-step instructions, is added as contextual knowledge to the ChatGPT Assistant, which is then integrated into the VR environment, allowing for an interactive AI-assisted operational experience with the juice mixer digital twin. \textbf{TODO: the components in the Figure 1 should focus more on what's happening, describe in steps, what is happening, make it more prominent, the libraries and other.}}
%\label{fig:architecture}
%\end{figure}

%%%
\vspace{-2mm}
\section{Conclusion and Future Work}
%We demonstrate an AI-powered immersive assistance system for supporting task training and execution in industrial settings. By employing a virtual juice mixer testbed, we showcase the potential of our system to reduce cognitive load and enhance operational efficiency. 

%In future developments, we aim to optimize the user interface based on user feedback and advance our system's contextual understanding by incorporating detailed tracking of the expert's actions and the user's environment within VR, leveraging GPT-4-vision\footnote{\url{https://platform.openai.com/docs/guides/vision}} technologies. We also plan to enrich the AI assistant's responsiveness to the user's situational context, potentially measuring physiological indicators to tailor support more precisely. Additionally, a comparative study on manual versus automated AI assistance invocation will inform our approach to enhancing immersive learning and operational support. These efforts will continue to refine our system, driving towards an adaptive, context-aware AI assistant that sets a new benchmark in industrial applications.

In this work, we presented an AI-powered immersive assistance system to interactively support users in task training and execution in industrial settings. Using a virtual juice mixer testbed, we demonstrated the potential of our system to enhance productivity and streamline complex operational tasks.

In the future, we will investigate ways to support users in a more precise and effective way.
For example, by examining how the user interface can impact the user behavior, or by incorporating physiological indicators.
%based on feedback and improve contextual understanding through detailed tracking of both expert actions and user environment within VR. 
Also, novel large language models, e.g., GPT-4-vision\footnote{\url{https://platform.openai.com/docs/guides/vision}}, will enable us to extract multimodal embeddings from the experts' video recordings, which could enhance the quality of the contextual information and thus, improve the precision of the assistants' guidance. 
%This will leverage GPT-4-vision\footnote{\url{https://platform.openai.com/docs/guides/vision}} technologies and physiological indicators for more precise support. Evaluating large language models will help refine video snippet retrieval based on multimodal embeddings. 
%A comparative study on manual vs. automated AI assistance invocation will also guide our approach to immersive learning and operational support. 
Finally, we plan to combine our data-driven AI approach with a theory-driven one, e.g., based on cognitive-inspired recommender systems~\cite{kowald2015refining,lex2021psychology}, to enhance the transparency and understandability of our AI-powered immersive assistant.
%With these efforts, we aim to develop a context-aware AI assistant that provides precisely contextualized information and sets a new benchmark for future research in interactive AI-powered task guidance.

\textbf{Acknowledgements.} This work was funded by the FFG COMET module Data-Driven Immersive Analytics (DDIA).

%%%%%%%%%%%%%%%%%%%%%%%%%%%%%%%%%%%%%%%%%%%%%%%%%%%%%%%%%%%%%%%%%%%%%%%%

%%% Use this environment to include acknowledgements (optional).
%%% This will be omitted in doubleblind mode.

% \begin{ack}
% This work was supported by the “DDIA” COMET Module within the COMET—Competence Centers for Excellent Technologies Programme, funded by the Austrian Federal Ministry for Transport, Innovation and Technology (bmvit), the Austrian Federal Ministry for Digital and Economic Affairs (bmdw), FFG, SFG, and partners from industry and academia. The COMET Programme is managed by FFG.
% \end{ack}

%%%%%%%%%%%%%%%%%%%%%%%%%%%%%%%%%%%%%%%%%%%%%%%%%%%%%%%%%%%%%%%%%%%%%%%%

%%% Use this command to include your bibliography file.

%\balance
\bibliography{mybibfile}

\end{document}